\documentclass[twocolumn, twoside,prl]{revtex4}

\usepackage{amsmath}
\usepackage{amssymb}
\usepackage{latexsym}
\usepackage{revsymb}
\usepackage{epsfig}

\newcommand{\barr}{\begin{eqnarray}}
\newcommand{\earr}{\end{eqnarray}}
\newcommand{\ra}{\rangle}
\newcommand{\la}{\langle}
\newcommand{\beq}{\begin{equation}}
\newcommand{\eeq}{\end{equation}}

\newcommand{\up}{\uparrow_x}
\newcommand{\dn}{\downarrow_x}
\begin{document}

\title{Each instant of time a new Universe}
\author{Yakir Aharonov$^{1,2}$}
\author{Sandu Popescu$^{3}$}
\author{Jeff Tollaksen$^2$}
\affiliation{$^1$ School of Physics and Astronomy, Tel Aviv University,  %
Tel Aviv, Israel}
\affiliation{$^2$ Chapman University, Schmid College of Sciences, Orange, CA 92866, USA}
\affiliation{$^3$ H.H.Wills Physics Laboratory, University of Bristol, %
 Tyndall Avenue, Bristol BS8 1TL, U.K.}

\begin{abstract}
We present an alternative view of quantum evolution in which each moment of time is viewed as a new "universe" and time evolution is given by correlations between them.
\end{abstract}

\date{}

\maketitle

\section{Introduction}

Since the dawn of civilization, mankind tried to understand the meaning of the inexorable flow of time. One of
the philosophical ideas, whose origins can be traced back to Heraclit from Efes, states that the universe gets
re-created again and again, at every instant of time. ``You never bathe twice in the same river" said Heraclit.
Every instant new water, every instant a new universe. This idea, of course, is different from the usual one in
which we view the universe as unique, and the objects which inhabit it as just changing their state in time.

Going from philosophy to physics, the way physics is presently formulated is based on the idea of unique
universe and evolving objects.  But is it possible to reformulate physics to incorporate the idea of a new
universe for each instant? As far as classical physics is concerned, the reformulation is rather trivial. We find
however that quantum mechanically things are more complicated.  The standard formalism of quantum mechanics
appears not to allow such a reformulation.  It turns out nevertheless that the reformulation is possible if we
use the two-states formalism\cite{ABL},\cite{multi-time-states}.

\section{Toy models: The difficulty}

In preparation for discussing the time evolution from this alternative point of view, let us start by asking a
much simpler question. Consider first classical mechanics. Suppose a particle evolves such that its trajectory
is $x(t)$ (fig.1). Consider now a set of $N$ time moments, $t_1$, $t_2$,..., $t_N$.  Is it possible to prepare
instead of this single particle a set of $N$ particles such that if we perform at some given time $\tau$
measurements on these $N$ particles we'll get the same information as we would have obtained by making measurements
at $t_1$, $t_2$...$t_N$ on the original single particle considered before?

The solution is quite simple: One has to prepare the $N$ particles (fig. 2) such that \barr
x_1(\tau)&=&x(t_1)\nonumber\\ x_2(\tau)&=&x(t_2)\nonumber\\&\vdots&\nonumber\\x_N(\tau)&=&x(t_N).\earr When $N$
increases and the time intervals $t_{i+1}-t_i$ decrease,  we could say that the $N$ particles lay down, at a
single moment of time (at $\tau$) the entire history of the original particle. One particle at N times is thus
equivalent to N particles at one time.
\begin{figure}
\epsfig{file=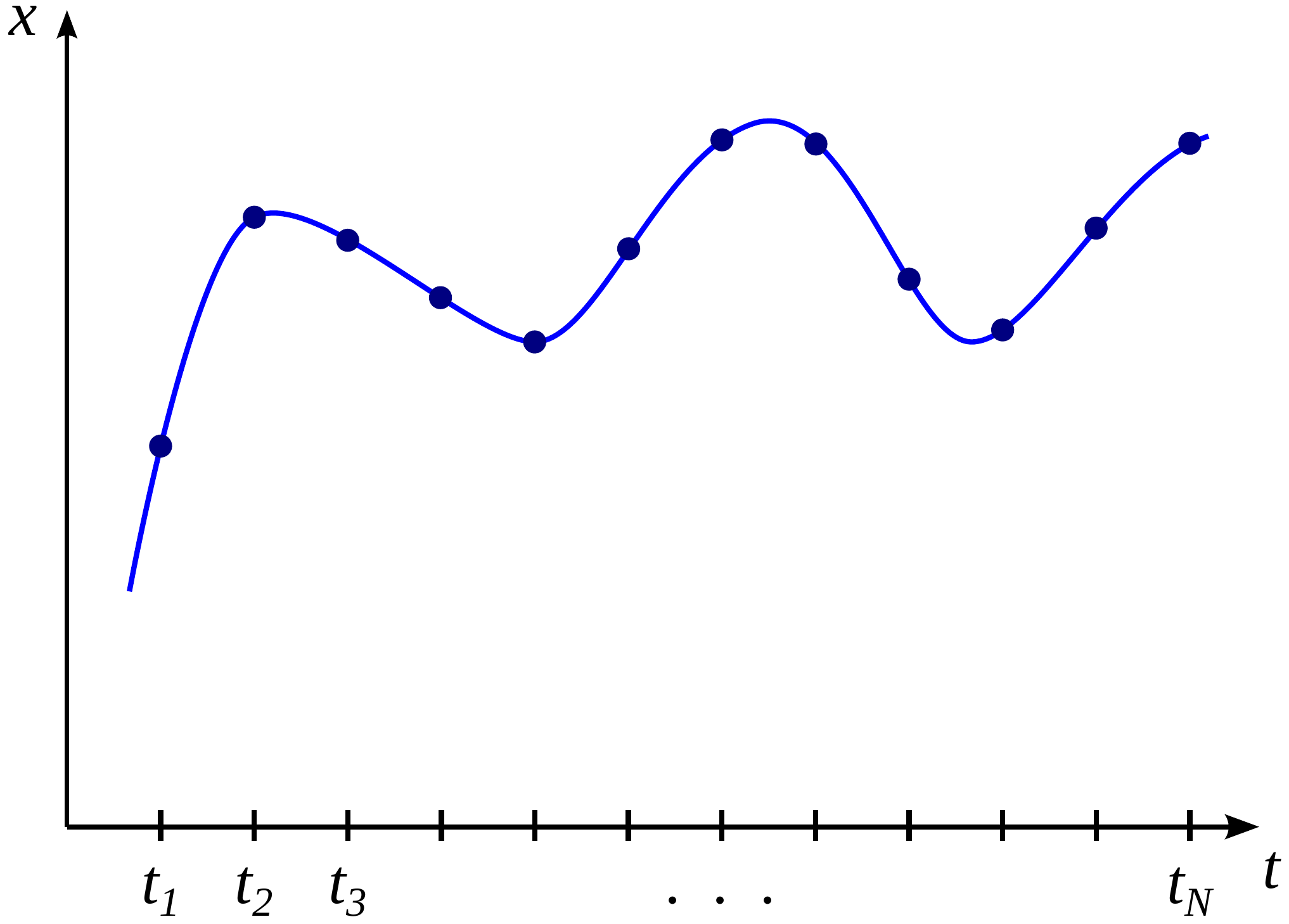, scale=0.35}\caption{A single classical particle at $N$ moments in time.}
\end{figure}

\begin{figure}
\epsfig{file=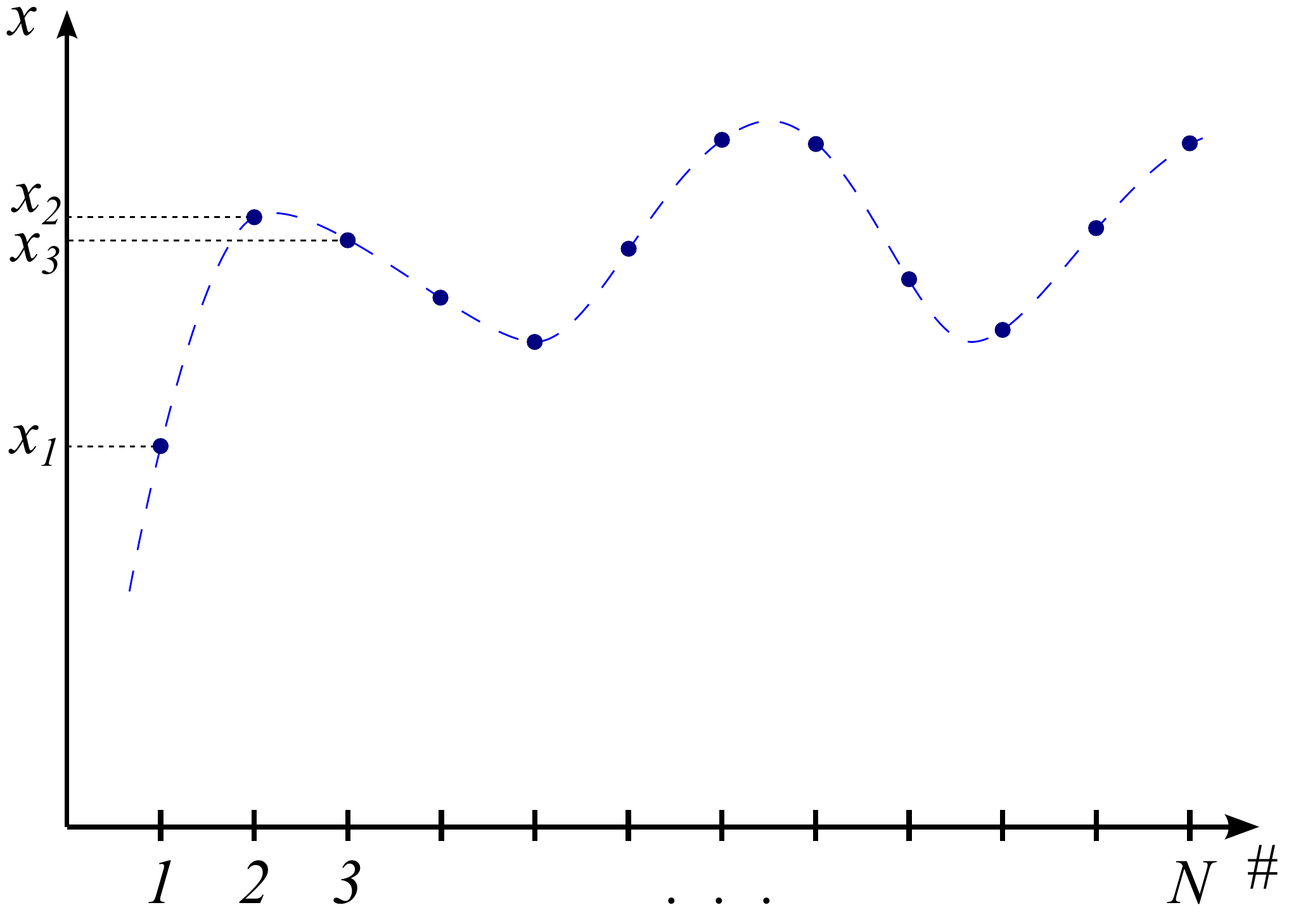, scale=0.35}
\caption{$N$ classical particles at a given time $\tau$. The position of particle $i$ is the same as the position of the original particle at time $t_i$ (see fig 1).}
\end{figure}
Can we do the same for a quantum mechanical particle?

Consider the simple case of a spin 1/2 particle, prepared in some initial state $|\psi\ra$ and having the
hamiltonian $H=0$. In this case the time evolution of the particle is trivial,

\beq |\psi(t)\ra=|\psi\ra.\label{evolution}\eeq Could we now prepare $N$ spin 1/2 particles such that if we
perform measurements on them at some time $\tau$ we get the same information as we could get by measuring the
state of the original particle at $N$ different time moments, $t_1$, $t_2$...$t_N$?  Since the state of the
original particle at all these moments is $|\psi\ra$, one would suppose that this task can be accomplished (fig.
3) by preparing the $N$ particles each in the same state $|\psi\ra$, that is

\beq |\Psi\ra_1|\Psi\ra_2...|\Psi\ra_N.\label{zspins}\eeq
\begin{figure}
\epsfig{file=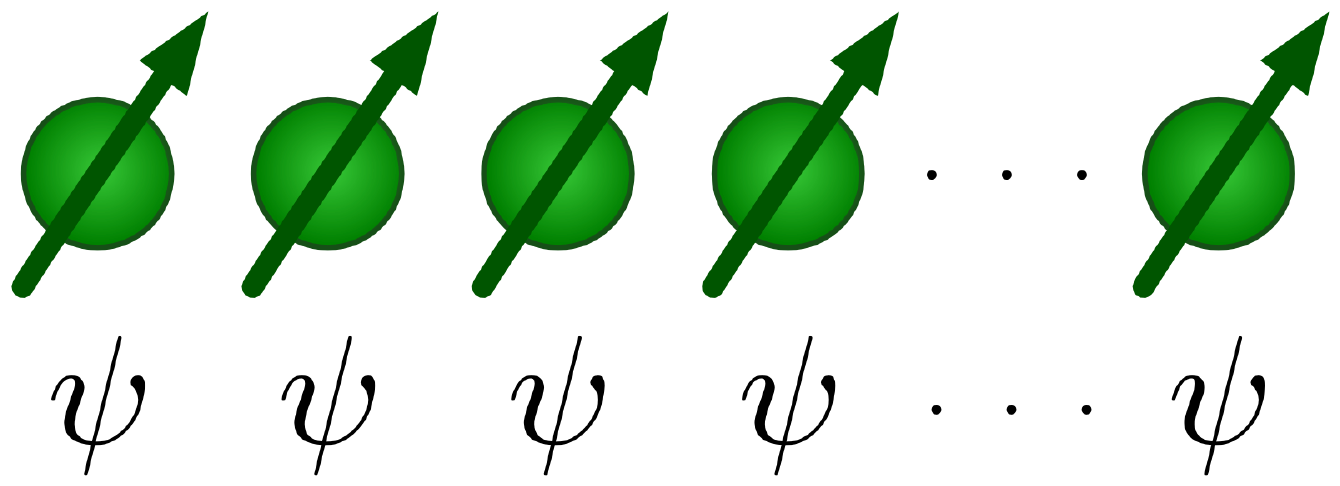, scale=0.35}
\caption{$N$ spin 1/2 particle at time $\tau$, each prepared in the state $\psi$. They do not constitute however an appropriate mapping of $N$ time moments of a single spin 1/2 particle}
\end{figure}

But this mapping is not appropriate for many reasons. First of all, the state \eqref{zspins} contains {\it too
much} information. Indeed, suppose somebody prepared the original particle in state $\psi$ and gave it to us without telling us what the state is. Then, as we have a single particle (i.e. a single copy of an unknown state $\psi$), there is no way of
learning what its state is. However, in \eqref{zspins} we have $N$ particles, all in the same state -  by
making different measurements on the different copies and looking at the statistics of the results we can learn
the state (better and better as $N$ becomes larger).

Second, the time evolution \eqref{evolution} contains subtle correlations, which usually are not noticed, and
which do not appear in the state \eqref{zspins}. Suppose, for example, that the state $|\psi\ra=|\sigma_z=1\ra$,
i.e. the spin is polarized ``up'' along the $z$ axis. It is generally considered that since the particle is at
every moment in a definite state of the $z$-spin component, the $z$-spin component is the only thing we know
with certainty about the particle - no other spin component commutes with $\sigma_z$ hence it is not
well-defined. However, there are {\it multi-time} variables whose values are known with certainty, given the
evolution \eqref{evolution}. For example, although the $x$ spin component is not well defined when the spin is
in the $|\sigma_z=1\ra$ state, we know that it is constant in time, since the hamiltonian is zero. Thus, in
particular, the two-time observable \beq\sigma_x(t_2)-\sigma_x(t_1)=0.\label{xcorrelations}\eeq As described in
\cite{aa}, \cite{multi-time-states}, this observable can be measured in the following way. Following von
Neuman's measuring formalism, consider a measuring device whose pointer position is denoted by $q$ and its
canonical conjugate momentum $p$ and let  the interaction between the spin and the measuring device be described
by the interaction hamiltonian \beq H_{int}=-\delta(t-t_1)p\sigma_x +\delta(t-t_2)p\sigma_x.\label{von_neumann}\eeq We also assume
that the Hamiltonian of the measuring device at all other times is zero (i.e. the pointer doesn't move by
itself). Due to the strong coupling between the spin and the measuring device during the measurement (the delta
function coupling) we can neglect during the measurement any other interaction affecting the spin. From the
Heisenberg equations of motion we obtain
\beq{{dq}\over{dt}}=i[q,H_{int}]=\bigl(\delta(t-t_2)-\delta(t-t_1)\bigr)\sigma_x(t). \label{vonNeuman}\eeq The
equation can be integrated easily, by noting that $\sigma_x$ doesn't change in the two (infinitezimally) short
interaction times ($t_1-\epsilon$ to $t_1+\epsilon$ and  $t_2-\epsilon$ to $t_2+\epsilon$) when the measuring
interaction takes place since it commutes with the interaction Hamiltonian. Hence, integrating \eqref{vonNeuman}
we obtain \beq q(t_2+\epsilon)-q(t_1-\epsilon)=\sigma_x(t_2)-\sigma_x(t_1),\eeq in other words, the difference
between the final and initial positions of the pointer indicates the value of the two-time observable
$\sigma_x(t_2)-\sigma_x(t_1)$. Crucially, this is a measurement which tells {\it only} the value of
$\sigma_x(t_2)-\sigma_x(t_1)$ but not the value of $\sigma_x(t_1)$ or $\sigma_x(t_2)$ separately. Now, when the
Hamiltonian acting on the spin is zero, $\sigma_x(t_1)=\sigma_x(t_2)$ and therefore the measuring device will
indicate \beq q(t_2+\epsilon)-q(t_1-\epsilon)=\sigma_x(t_2)-\sigma_x(t_1)=0.\eeq

It is also important to note that, given that the spin Hamiltonian is zero, after the measurement is finished,
i.e. after $t_2$, the spin is brought back in the initial state, regardless of what this state is. Indeed,
suppose that the initial state of the spin is $|\psi\ra$. By decomposing $|\psi\ra$ in the $x$-basis, in the
Schrodinger representation the evolution of the spin and measuring device is given by \barr
&&|\psi\ra|q=0\ra=(\alpha|\up\ra+\beta|\dn\ra)|q=0\ra\nonumber\\
&\rightarrow&\alpha|\up\ra|q=-1\ra+\beta|\dn\ra|q=1\ra\\
&\rightarrow&\alpha|\up\ra|q=0\ra+\beta|\dn\ra|q=0\ra= |\Psi\ra|q=0\ra\nonumber\earr where the
two arrows describe the evolution of the spin and measuring device at at $t_1$ and $t_2$ respectively.  (Note
that the first interaction shifts the pointer with the value $-\sigma_x$ while the second shifts the pointer
with $+\sigma_x$.)

Coming back to the $N$ spins in the state \eqref{zspins}, there is no correlation whatsoever in between the $x$
components of, say, particles 1 and 2 which were intended to describe the original particle at times $t_1$ and
$t_2$. More over, it is not only the $x$ spin component for which the original particle presents such time
correlations, but all spin components. That is, \eqref{xcorrelations} generalizes to

\beq \sigma_{\hat n}(t_1)=\sigma_{\hat n}(t_2)=...=\sigma_{\hat n}(t_N)\label{directionn}\eeq for any direction
$\hat n$. (Following the above
procedure, we would then obtain $\sigma_{\hat n}(t_i)-\sigma_{\hat n}(t_j)=0$ for any $i$ and $j$.) Obviously, the $\hat n$ spin components of the $N$ spins (except for the $z$ components) are not
correlated in this way.

We reached thus the conclusion that the $N$ particles in the state \eqref{zspins} {\it do not} describe
faithfully the behavior of the original particle at $t_1$...$t_N$. The question is now whether there exists any
state of $N$ particles that could be such a faithful representation? It is easy to see that the answer is ``no".
Indeed, there is no state of $N$ spins such that \beq \sigma_{\hat n}^1=\sigma_{\hat n}^2=...=\sigma_{\hat
n}^N\label{correlations}\eeq for every direction $\hat n$. At best, one may find a two particle state - the
singlet state- for which the spins are {\it anti-correlated} instead of correlated i.e. \beq \sigma_{\hat
n}^1=-\sigma_{\hat n}^2\eeq for every $\hat n$. And even anti-correlation cannot be extended to more than two
particles. It appears thus that we reached a dead end.

It is tempting to think that the reason we  arrived at this dead end is that we didn't take into account that
quantum mechanically measurements modify the state of the measured system. We could say that the state of the
original particle is constant in time, \eqref{evolution}, and thus it is modeled by the $N$ spins in state
\eqref{zspins} {\it only} if no measurements are performed. If measurements are performed, the time evolution of
the original particle is no longer given by \eqref{evolution}, so we shouldn't expect to model it by
\eqref{zspins}. But this is actually {\it not} the true reason for our difficulties. Indeed, even if we don't
actually measure $\sigma_x(t_2)-\sigma_x(t_1)$ but merely {\it compute} it for the evolution \eqref{evolution},
we see that it has not the same value as if we compute it for the state \eqref{zspins}. So problems arise
already at this stage - the state \eqref{zspins} simply doesn't contain the correlations which the free
evolution of the original particle prescribes.

\section{Toy models: The solution}

We now arrived at a crucial point. Although a state of $N$ spin 1/2 particles with complete correlations among
all their spin components as required by \eqref{correlations} doesn't exist in the usual sense, there exist pre-
and post-selected states \cite{multi-time-states} with this property. As we show now, $N$ spin 1/2 particles in
a suitably prepared pre- and post-selected state can, at one time, $\tau$, mimic $N$ moments of time in the
evolution of a single spin 1/2 particle.

The procedure we will use has three steps. At time $\tau-\epsilon$ we prepare an initial state. At time $\tau$
we perform the measurements that are supposed to simulate the evolution of the original single spin. At time
$\tau+\epsilon$ we perform an additional measurement; only if this measurement is successful, we deem
our simulation procedure to have succeeded. Specifically, consider $2N-1$ spin 1/2 particles. $N$ of them will
be used as "spins", and we denote them by $S_0$,$S_1$...$S_N$. They will simulate $N$ time moments of the evolution of our original spin. For example, a measurement at time $\tau$
on the spin $S_k$ should simulate the measurement on the original spin at time $t=t_k$. Furthermore a two time
measurement on the original spin, say at $t_k$ and $t_l$ will be simulated by a measurement on spins $S_k$ and
$S_l$ and so on. The other $N-1$ spins are ancillas, which we denote by $A_1$,$A_2$...$A_N$. They are used for
helping in our procedure, however, no measurements will be performed on them at $\tau$. We arrange the "spins" and ancillas as
illustrated in fig4.

\begin{figure}
\epsfig{file=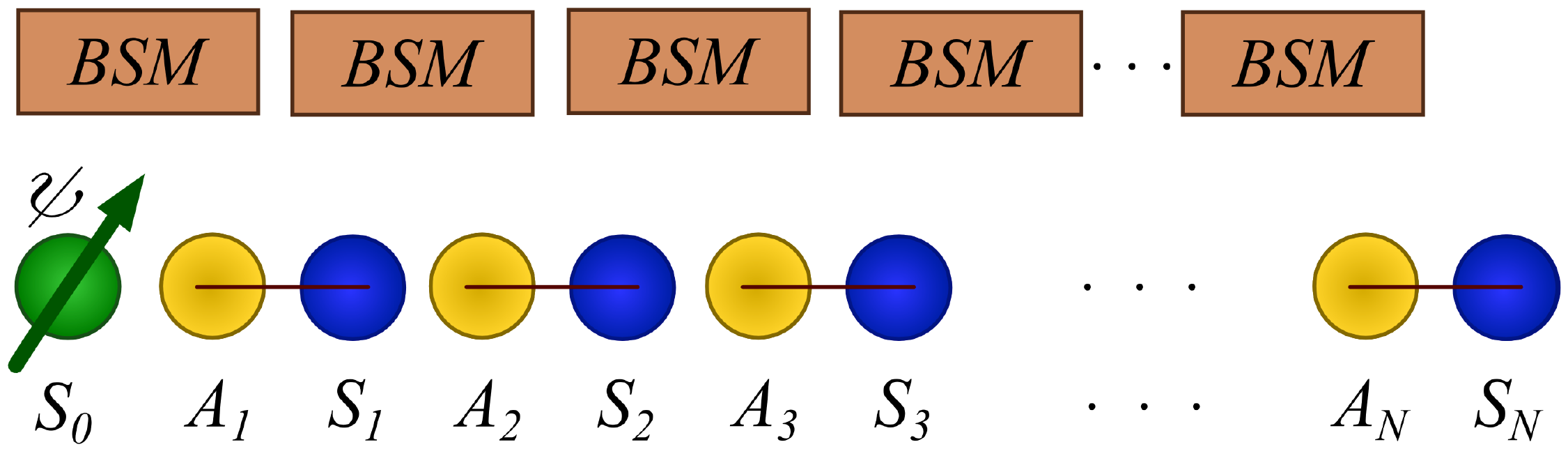, scale=0.35}
\caption{$N+1$ "spins" and $N$ ancillas. At time $\tau-\epsilon$ spin $S_0$ is prepared in state $\psi$ while ancilla $A_k$ is maximally entangled with spin $S_k$ (the maximal entanglement is represented by the continuous line connecting the ancilla with the spin.) At time $\tau+\epsilon$ a "Bell state" measurement (BSM) is performed on spin $S_{k-1}$ and ancilla $A_k$. The experiment is deemed successful if and only if each of the BSMs yields the outcome corresponding to the maximally entangled state $|\Phi\ra_{S_{k-1},A_k}={1\over{\sqrt2}}\sum_{i=0}^1 |i\ra_{S_{k-1}}|i\ra_{A_k}$. In case of a successful experiment, if measurements were performed on spins $S_0$...$S_{N-1}$ at time $\tau$, their results simulate measurements performed at $N$ moments of time on a single spin prepared in state $\psi$ and evolving with a Hamiltonian equal to zero.}
\end{figure}

 At $\tau-\epsilon$ we prepare the particles in the initial state  \beq |\psi\ra_{S_0}
|\Phi\ra_{A_1, S_1}\cdots|\Phi\ra_{A_N, S_N}\eeq where $|\Phi\ra$ is a maximally entangled state of the ancilla
$A_k$ and the associated spin, $S_k$, namely $|\Phi\ra_{A_k,S_k}={1\over{\sqrt2}}\sum_{i=0}^1
|i\ra_{A_k}|i\ra_{S_k}$ and where $|i\ra$, $i=0,1$ represent some arbitrary base vectors. At a later time,
$\tau+\epsilon$ we subject the pairs of spins composed by spin $S_{k-1}$ and ancilla $A_k$ to a
measurement of an operator that has the maximally entangled state
$|\Phi\ra_{S_{k-1},A_k}={1\over{\sqrt2}}\sum_{i=0}^1 |i\ra_{S_{k-1}}|i\ra_{A_k}$ as one of its nondegenerated
eigenstates (for example the "Bell" operator). Now, as it is easy to directly verify, in the case when all these measurements performed at
$\tau+\epsilon$ yield the outcome corresponding to this state, then measurements performed at $\tau$ on the
"spins" reproduce the same statistics as measurements on the original single spin.

The above procedure may seem rather convoluted. However, it has a very simple interpretation in the language of
pre- and post selected states \cite{aa, multi-time-states}. The procedure simply prepared a particular pre- and
post-selected state of the $N$ spins $S_1$...$S_N$. The defining characteristic of this state is that
the post-selected state of one particle is completely correlated to the pre-selected state of the next particle
as illustrated in fig. 3. Technically this is possible because post-selected states propagate backwards in time
and behave as complex conjugates of pre-selected states.  This accounts for the correlations that cannot be
created when we consider a pre-selected only state (i.e. a state prepared at $\tau-\epsilon$).

The idea of pre- and post-selected states and the above way of preparing them was discussed in detail in
\cite{multi-time-states}. Using the notation $ {\bf \Phi}_{k+1,k}^{\tau_-,\tau_+}$ for the {\it maximally
entangled} two-time state \beq{\bf \Phi}_{k+1,k}^{\tau_-,\tau_+}=\sum_i
|i\ra_{S_{k+1}}^{\tau_-}~{}_{S_k}^{\tau_+}\la i|\label{twotime}\eeq where $\tau_\pm=\tau\pm\epsilon$, from the
results in \cite{multi-time-states} it follows that in our case the pre-and post-selected state of the
spins is \beq \label{state}{\bf \Phi}_{N,N-l}^{\tau_-,\tau_+}\cdots{\bf
\Phi}_{2,1}^{\tau_-,\tau_+}{\bf \Phi}_{1,0}^{\tau_-,\tau_+}|\Psi\ra_{S_0}^{\tau_-}.\eeq Note that this pre-and
post-selected state refers solely to the spins; the ancillas were there only to help prepare this
state.

\begin{figure}
\epsfig{file=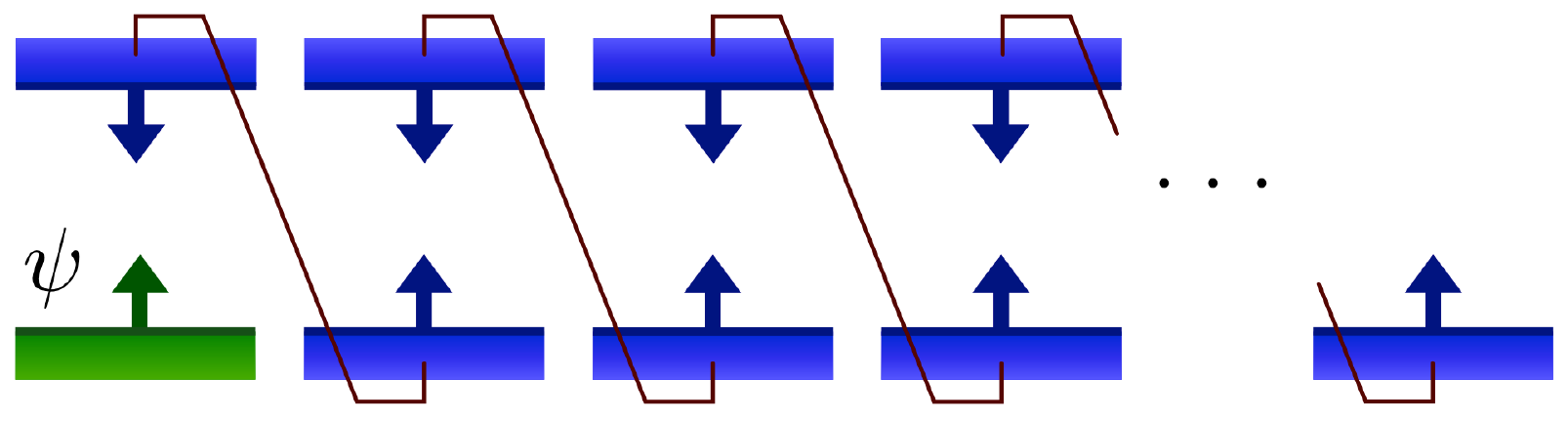, scale=0.35}
\caption{Correlation between the spins $S_0$...$S_N$ .
The post-selected state of one particle is correlated
to the pre-selected state of the next particle. Upward arrows denote "pre-selected" state, which point towards the future while downward arrows denote "post-selected" states that point towards the past.}
\end{figure}

The pre- and post-selected state \eqref{state} explicitly shows the two main characteristics of the time
evolution of the original spin. On one hand, at the initial time $t_0$ the original spin was prepared in the
state $|\Psi\ra$. To this corresponds the fact that in \eqref{state} the pre-selected state of $S_0$ is the same
as that of the original spin. On the other hand, we also know \eqref{directionn} that the spin components
along any direction, although undefined, are constant in time, that is, they are fully correlated. These
correlations are realised in \eqref{state} via the complete correlations between the post-selected state of one
spin and the pre-selected state of the next \eqref{twotime}. (Indeed, it is easy to verify that the maximally
entangled state \eqref{twotime} is invariant under a simultaneous change of basis for $S_{k+1}$ and $S_k$.)

\section{Every moment of time a new universe}

Up to this point we only dealt with a toy model. We now come to the main question, namely how to interpret the
time evolution of a quantum particle from the point of view of the philosophical idea of ``each moment of time a
new universe''. As far as classical physics is concerned, we could formalize this idea by associating a separate
configuration-space to each moment of time. A moving particle would then correspond to one particle in each
space, having their positions appropriately correlated. In effect, this would mean associating a different
configuration space to each of the $N$ particles in fig 1b and ``stacking" them one on top of the other along
the time axis, see fig?. Naively, one would expect that quantum mechanically this would correspond to
associating to each moment of time a separate Hilbert space. The total Hilbert space would be therefore ${\cal
H}= {\cal H}_0\otimes{\cal H}_1\otimes\cdots\otimes{\cal H}_N$. The problem however, as we saw before, is that
no state in such a Hilbert space can account for the desired correlations. The solution follows from the above
described mapping between the time evolution of the spin 1/2 particle and $N$ spin 1/2 particles: we have to
associate {\it two} Hilbert spaces to each moment of time and``stack'' them on top of each other along the time
axis (fig. 6a).

\begin{figure}
\epsfig{file=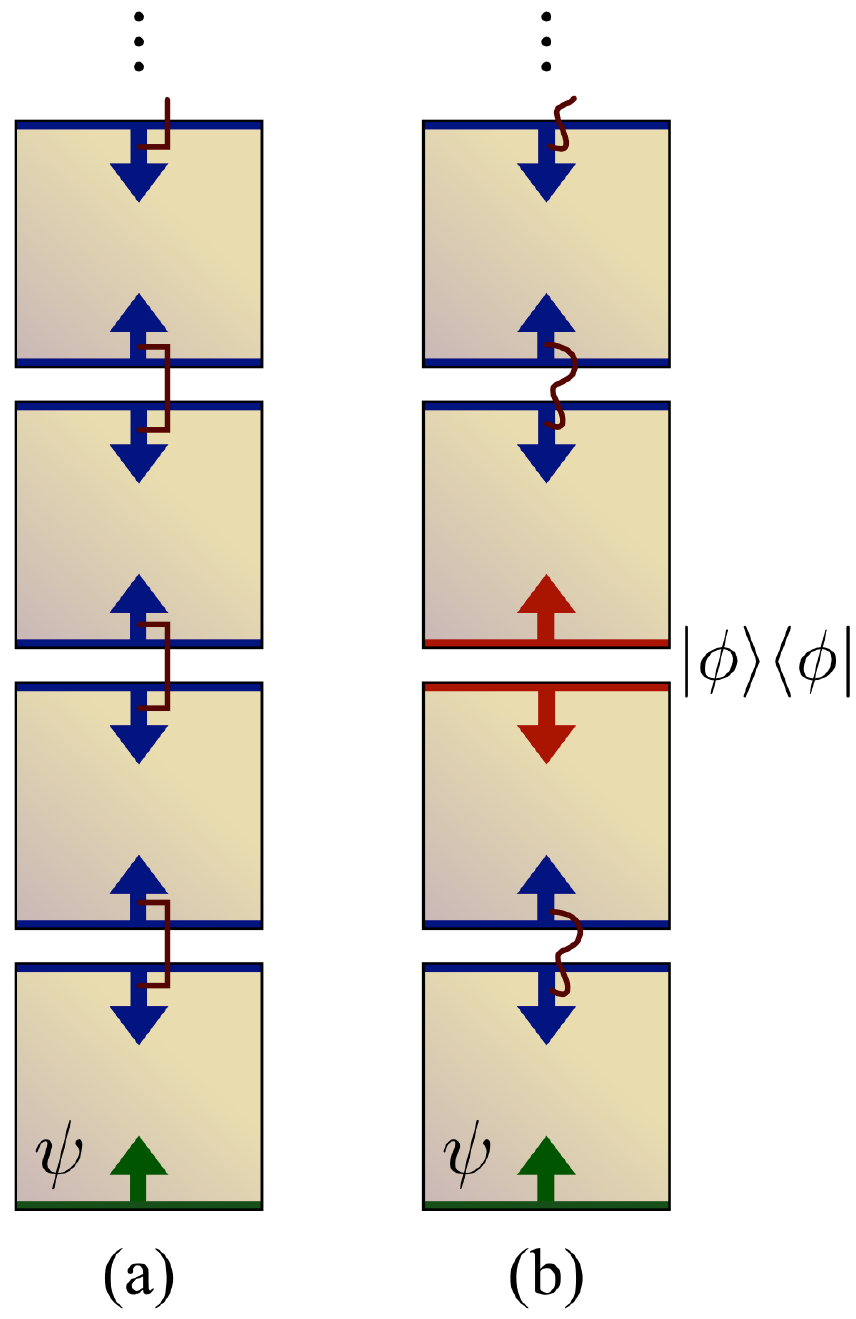, scale=0.4}
\caption{(a) Each moment of time is a little "brick". The Hilbert space at the future boundary of one time moment is maximally entangled - with complete correlation - with the Hilbert space at the past boundary of the next time moment (straight line). The state $\psi$ is associated only to the moment $t_0$ where it was prepared. (b) A more general time evolution. A measurement with a collapse on state $\phi$ disentangles the two subsequent moment of time. Non-trivial unitary time evolution at all other time is represented by maximal entanglement but between appropriately rotated bases (squiggled line).}
\end{figure}

We consider time as being discrete, made out of finite time intervals, stacked one on top of the other
like little bricks. To each moment of time we associate two Hilbert spaces: a Hilbert space of ket vectors at
the time boundary towards the past and a Hilbert space of bra vectors at the time boundary towards the future.
The total Hilbert space is therefore of the form ${\cal H}_N\otimes\cdots\otimes{\cal H}_1^{\dagger}\otimes{\cal
H}_1\otimes{\cal H}_0^{\dagger}\otimes{\cal H}_0$. The state of a quantum system during each
instant of time is thus determined by two wave-functions. One of them is fixed at the past time boundary of the
time interval and is ``evolving" towards the future, the other one is fixed at the future boundary of the time
interval and it is ``evolving" towards the past. We shall call these two states ``pre-selected" and
``post-selected".

Consider now the example we presented at the beginning of this paper: At time $t_0$ the spin was prepared in state $|\Psi\ra$ and then the evolution is trivial, i.e. the Hamiltonian is zero. In our new formalism the preparation at time $t_0$ corresponds having at time $t_0$ the forward evolving state $|\Psi\ra_{t_0}\in H_0$. At all other moments of time we know that all spin components are completely correlated. We describe this by
taking the wavefunctions at the common boundary of two subsequent moments (i.e. the post-selected state of the earlier
moment  and the pre-selected state of the later moment) to be {\it maximally entangled} and
{\it completely correlated}:  $ \sum_i|i\ra_{t_1}{}_{t_0}\la i|\label{identity}$  and so all at all times.  Here the
vectors $|i\ra_{t_1}$ and ${}_{t_0}\la i|$ form arbitrary orthonormal bases in  ${\cal H}_1$ and ${\cal H}_0^{\dagger}$
respectively and represent the same physical state (such as $|\sigma_z=1\ra$ and $\la \sigma_z=1|$). Since all time moments are correlated this way, the time evolution now looks like a chain
of connected time intervals. Hence the total state is
\beq \sum_k|k\ra_{t_N}{}_{t_{N-1}}\la k|\cdots\sum_i|j\ra_{t_2}{}_{t_1}\la j|\sum_i|i\ra_{t_1}{}_{t_0}\la i|~|\Psi\ra_{t_0}.\eeq

\bigskip

The example above can be easily generalized to arbitrary quantum evolutions which consist of unitary evolutions and measurement induced collapses (fig 6b). When the hamiltonian is non-zero
subsequent moments of time continue to be maximally entangled, but the correlation is now ``skewed". That is,
the correlation between two moments of time is now of the form

\beq \sum_i |u_i\ra_{t_2}{}_{t_1}\la i|\label{unitary}\eeq where \beq
|u_i\ra_{t_2}=U_{2,1}|i\ra_{t_2}.\eeq Here $U_{2,1}$ represents a unitary transformation acting on ${\cal
H}_2$, the pre-selected Hilbert space at time $t_2$, numerically equal to $U(t_2, t_1)$, the unitary that
describes the evolution of the particle from $t_1$ to $t_2$ in the standard quantum description.

Note that, similarly to  $ \sum_i|i\ra_{t_1}{}_{t_0}\la i|\label{identity}$ the state \eqref{unitary} is also a maximally entangled state
between the two moments of time, and also leads to full correlations. The only difference is that the
correlations are now not between an arbitrary vector ${}_{t_1}\la \xi|$  and the corresponding vector
$|\xi\ra_{t_2}$ but between ${}_{t_1}\la \xi|$ and $U_{2,1}|\xi\ra_{t_2}$.

On the other hand when a measurement occurs it completely disturbs all the observables that do not commute with the measured observable - their values before the measurement are no longer correlated with their values after the measurement. At the same time a collapse means the introduction of a new boundary condition for times following the collapse as well as a post-selection for times previous to the collapse. Suppose an instantaneous ideal von Neuman measurement took place at time $t_k$ and let $|\phi\ra$ be the eigenstate corresponding to the observed eigenvalue. In our alternative formalism this collapse is implemented simply by $|\phi\ra_{t_{k+1}}{}_{t_k}\la\phi|$.

We have now reached our desired alternative description of time evolution. To summarize:
\begin{itemize}
\item Each moment of time is indeed one "Universe" but it has associated to it not one but two Hilbert spaces, one corresponding to the "past" boundary of this time moment and one to its "future" boundary
\item Time evolution is represented by correlations between subsequent moments of time, more precisely between the "future" boundary of the earlier time moment and the "past" boundary of the later time moment.
\item Unitary time evolution is implemented by maximal entanglement between subsequent moments of time.
\item A measurement induced collapse destroys the entanglement and effectively decouples the entire time evolution up to that moment by what happens later; technically, the state before the moment of collapse is in a direct product with the state after it.
\item  A partial collapse, such as one due to an incomplete measurement will result inentanglement but  less than maximal.

\end{itemize}

\section{Probabilities}

In the above section we described the time evolution of a quantum system in the "each moment of time a new universe" paradigm. To complete our description there is one more item to address: how to calculate probabilkities for the different events.

 The probability for a particular history to happen can be computed from the "history" state in a straightforward way, with one subtlety.  As it stands now, the history is open ended, both towards the remote past and the remote future; without puting boundary conditions at those two end, nothing can be said about the overall porobability of the history. The standard experimental questions however remove the need for these "cosmological" implications by completely separating a piece of time from its past and future by making complete measurements. The most well-known case is preparing at initial time $t_0$ the system in some state $|\Psi\ra$ and then performing a measurement at the final time $t_f$ and asking what is the probability to find the system in state $|\phi\ra$. Looking at our description above, we see that indeed these two measurements cut out a piece of time, i.e. the state starting at $t_1$ and ending at $t_2$ is completely disentangled from the rest. For example, if the Hamiltonian is zero, this piece is:

\beq {}_{t_f}\la \phi| \sum_k|k\ra_{t_N}{}_{t_{N-1}}\la k|\cdots\sum_i|j\ra_{t_2}{}_{t_1}\la j|\sum_i|i\ra_{t_1}{}_{t_0}\la i|~|\Psi\ra_{t_0}.\eeq

To find the probability of obtaining  for  $|\phi\ra$ when the system was prepared in $|\Psi\ra$ all we do is the following: for each moment of time we contract the vectors belonging to the past and future boundary conditions, i.e. we take the scalar product between the bra and ket vectors corresponding to the same time $t$. The result is the scalar product $\la\phi|\Psi\ra$; the absolute value of this, $|\la \phi|\Psi\ra|^2$  is the probability we are loong for. It is also immediate to show that in case the Hamiltonian is non-zero, our recepy leads to  $|\la \phi|U(t_f,t_0)|\Psi\ra|^2$, the expected result.

\section{Discussion}

We would like to emphasize what is arguably the most important property of this alternative view of time
evolution, namely that there is a complete harmony in between the actions to which the system is subjected and
their representation: If a measurement and its associated collapse occur at time $t$ it is {\it there and only there} that this information is present - the state $\Phi$ prepared by the collapse appears only at this time. When there are more measurements, each measurement and the state associated to its outcome appear at the time of measurement and only there. At other times, when no measurement is performed all we know is that the time correlations are preserved, and this is what our formalism shows.

Our description is in stark contrast with the usual one in which once we prepare a system in a state $|\psi\ra$ and we leave it undisturbed than at every subsequent moment of time the state continues to be $|\psi\ra$. As far as we are concerned however, the state $|\psi\ra$ doesn't characterize {\it directly} any other moment of time except when it was prepared; it does influence the physics at these other moments, but it does so only indirectly, via a chain of time correlations. What does directly characterize a time when no measurement is performed is that it is an unbroken link in a chain of correlations, nothing more than this; what propagates along the chain is a completely independent issue.

It is very interesting to ponder more carefully on the difference between a measurement and a unitary evolution from our point of view. What we see is a certain {\it complementarity} between kinematics and dynamics. When a measurement is performed we know the state at each of the two subsequent moments of time when the measurement took place:

\beq |\phi\ra_{t_{k+1}}{}_{t_k}\la\phi|.\eeq
On the other hand, when a unitary evolution takes place, the state at each of the two subsequent moments of time is completely uncertain, the state at one moment being entangled with the state at the next moment
\beq \sum_i |u_i\ra_{t_{k+1}}{}_{t_k}\la i|.\eeq

Furthermore, we note that every measurement is effectively an uncertain time evolution. This is a fact that, as far as we know, it is very rarely mentioned in discussions about quantum measurements. Yet, it is quite obvious. Indeed, as is well known all the observables that do not commute with the measured one are randomized up to some extent, hence their Heisenberg equations of motion must show an uncertain evolution. In its turn, this due to the fact that during the measurement the hamiltonian of the system is uncertain. Indeed, in the standard von Neumann measurement formalism (as used above in (\ref{von_neumann})) in order to measure an observable $A$ and register its value in the indication $q$ of a pointer we use an interaction Hamiltonian of the form
\beq H_{int}=\delta(t)Ap\eeq
where $p$ is the canonical momentum conjugate to the position $q$ of the pointer. Since the initial state of the pointer is well defined, say $|q=0\ra$, the momentum of the pointer, $p$ has a large uncertainty $\Delta p=\infty$. In its turn, since $p$ enters the interaction Hamiltonian, it means that as far as the system is concerned, its Hamiltonian is uncertain during the measurement.

The above observations, although not very commonly known, are nevertheless rather straightforward. What our new formalism shows however, is something more subtle: although a measurement is equivalent to an uncertain evolution, the collapse on a particular eigenstate of the measured observable is equivalent to a well-defined {\it superpositon of different time evolutions}\cite{superposition_evolutions}. Indeed, take for example a measurement of the $\sigma_x$, the $x$ component of the spin of a spin 1/2 particle performed at $t_k$. Suppose we found $\sigma_x=1$. According to our formalism the quantum state at the time of measurement is
\beq |\up\ra_{t_{k+1}}{}_{t_k}\la\up|.\eeq
This can be viewed as the superposition of two unitary time evolutions

\barr && |\up\ra_{t_{k+1}}{}_{t_k}\la\up|=\nonumber\\&&{1\over{\sqrt2}}\Big( |\up\ra_{t_{k+1}}{}_{t_k}\la\up|+|\dn\ra_{t_{k+1}}{}_{t_k}\la\dn|\Big)+\nonumber\\&&{1\over{\sqrt2}}\Big( |\up\ra_{t_{k+1}}{}_{t_k}\la\up|-|\dn\ra_{t_{k+1}}{}_{t_k}\la\dn|\Big).\nonumber\\
&&{}\earr

Hence our picture suggests a new kind of complementarity between having information about the state of a system versus having information about the dynamics: if one does not know the state, then our picture describes the dynamics as complete correlation.  If one obtains information about the state, then the multi-time correlations  between conjugate operators are made uncertain, i.e. one loses information about the dynamics in that interval of time. And as for a proper conjugacy relationship, there is a continuous graduation between the extremes. To see this complementarity, consider a partial measurement of $\sigma_x$ in which the measuring device gives the correct answer (i.e. $\sigma_x=\pm1$) with probability $|\alpha|^2$ and the wrong answer with probability $|\beta|^2$ and does this in a way which minimizes the disturbance to the state. This is obtained when the measuring device interacts with the spin via the unitary evolution

\barr &&|\up\ra|0\ra_M\rightarrow|\up\ra(\alpha|1\ra_M+\beta|-1\ra_M)\nonumber\\ &&|\dn\ra|0\ra_M\rightarrow |\dn\ra(\alpha|-1\ra_M+\beta|1\ra_M)
\earr
where $|0\ra_M$, $|1\ra_M$ and $|-1\ra_M$ are different states of the measuring device.
Obtaining the value $+1$ corresponds in our picture to partially destroying the complete correlation between the moments when the measurement occurred and leading to only non-maximal correlations\footnote{In the standard language this outcome corresponds to the Krauss operator $\alpha |\up\rangle\la\up| +\beta |\dn\rangle\langle \dn|$} :

\beq
\alpha |\up\rangle_{t_2}{}_{t_1}\la\up| +\beta |\dn\rangle_{t_2}{}_{t_1}\langle \dn|,
\label{e122}.
\eeq
We see that in the case $\alpha =\beta$, we have complete correlation, thus modeling the dynamics.  As $\alpha \rightarrow 1$ and $\beta \rightarrow 0$
 we obtain more and more knowledge about the state, while the entanglement, i.e. the dynamics, becomes more and more uncertain.

\section{Measurements on EPR states}

It is very interesting to analyze using our point of view the time evolution of two spin 1/2 particles in a singlet state. The evolution is illustrated in fig 7. At $t_0$ the two particles $A$ and $B$ are prepared in the singlet state

 \beq|S\ra_{AB, t_0}={1\over{\sqrt2}} |\up\ra_{A,t_0}|\dn\ra_{B,t_0}-{1\over{\sqrt2}}|\dn\ra_{A,t_0}|\up\ra_{B,t_0}.\eeq Then each particle evolves separately. That is, the time moments describing particle $A$ are maximally entangled with each other and the time moments describing the evolution of $B$ are maximally entangled with each other. There is however no entanglement between particles $A$ and $B$ at any other subsequent moment. This situation continues until we disturb the particles.

\begin{figure}
\epsfig{file=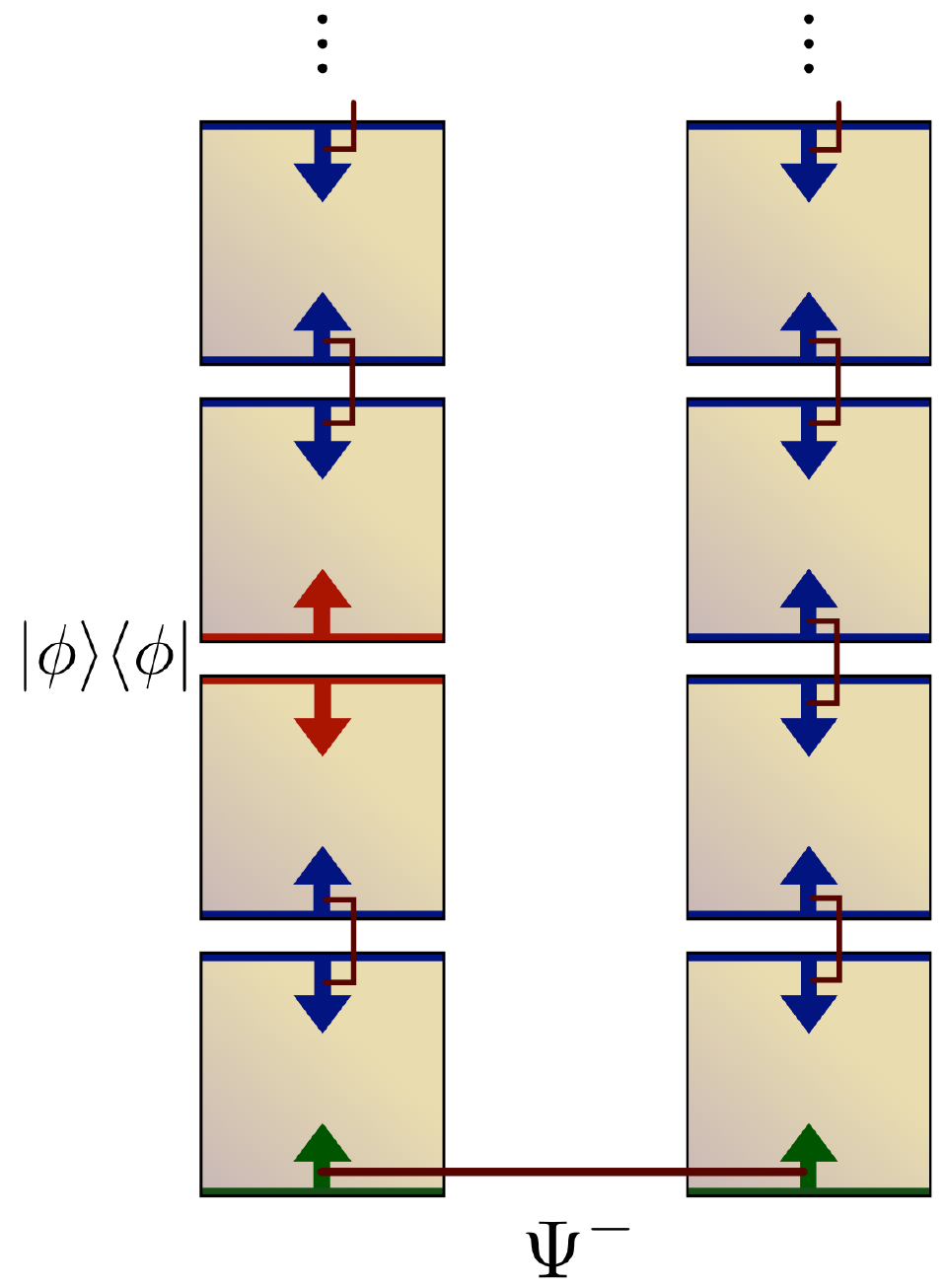, scale=0.35}
\caption{Two entangled spin 1/2 particles. Entanglement characterizes solely time $t_0$ where entanglement is produced. All other times are characterized by trivial time evolution, i.e. maximal entanglement between subsequent moments of time; there is however no entanglement between the particles associated to these times. Alice's measurement disentangles the time moments of her particle but have no effect on Bob's particle.}
\end{figure}

Suppose now that  at time $T$ Alice performs a measurement on particle $A$. For example, suppose she measures $\sigma_x^A$ and finds the value $+1$. According to our point of view, for particle $A$ the entanglement between the moments of time $T$ and $T+\epsilon$ is broken and in its place we have a direct product state $|\up\ra_{A,T+\epsilon}~{}_{A,T}\la \up|$. On the other hand, nothing happens to particle $B$ - its time moments continue to be maximally entangled.

This conclusion seems to contradict the standard quantum mechanical description. Indeed, in the usual description the state is $|S\ra_{AB}$ from time $t_0$ until time $T$ and at time $T$ the state collapses into the direct product state $|\up\ra_A|\dn\ra_B$. In particular, the collapse is symmetric with respect to who produces it: The time evolution is the same whether Alice measured $\sigma_x^A$ and found $+1$ or Bob measured $\sigma_x^B$ and found $-1$. In our description however, if Alice performs the measurement, the time evolution of her particle is affected and nothing happens to Bob's, while the opposite would be true if Bob were to perform the measurement.
One can check directly however that all {\it observational} consequences, i.e. the probabilities for all measurements, are the same in both descriptions. Our point of view however has two main advantages.

First of all, it is relativistically covariant {\it at the level of states}. Of course, both views are  relativistically covariant at the level of observed results. The standard way however is {\it non-covariant} as far as the state description is concerned. Indeed, the collapse occurs both at Alice and at Bob at time $T$, i.e. simultaneously in the reference frame in which we chose to work. Had we chosen a different reference frame, the moment at which the collapse occurs  for Bob's particle could have been different. On the other hand, in our description, nothing happens to Bob's particle when Alice performs a measurement, so no covariance problems arise.

We would like to emphasize however that the relativistic covariance at the level of wave-functions does not necessarily require to consider each moment of time a new universe; it is already present in a simpler version of time evolution, with a ``single universe'' but with two wave-functions, one propagating forward and the other backward in time \cite{quantum-relativistic}.

A second interesting feature of our description is that it makes it clear that the evolution of Alice's particle is {\it different} from Bob's particle, while in the standard description they looked the same. Indeed, in the standard description they appear symmetric - they are in a singlet until time $T$ and then they collapse together on a direct-product state. In our description is clear that for particle $A$ the time moments before and after $T$ are not maximally entangled while for particle $B$ they are. This difference could be checked if in addition to the measurement at time $T$ Alice also measures a two-time variable, say $\sigma_z^A(t_1)-\sigma_z^A(t_2)$ for $t_0<t_1<T<t_2$. Since the spin components along the $z$ direction before and after $T$ are not correlated, Alice could obtain +2, 0 or -2. On the other hand, if Bob were to measure $\sigma_z^B(t_1)-\sigma_z^B(t_2)$ he would obtain with certainty the value 0.

\section{Extensions}
Many more interesting situations are possible. An amusing one is illustrated in fig8. Here every moment of time is fully correlated with the second next. If effect this particle has a ``double life" - the even time moments describe a particle whose time evolution is $|\psi(t)\ra=|\psi_1\ra$ and the odd moments describe a particle whose time evolution is $|\psi(t)\ra=|\psi_2\ra$. As long as we do not take any action to connect them, such as a two-time measurement involving an odd and an even moment, the two lives of this particle do not interact with each other. It is interesting to speculate if such things exist in nature, and what their meaning would be.

\begin{figure}
\epsfig{file=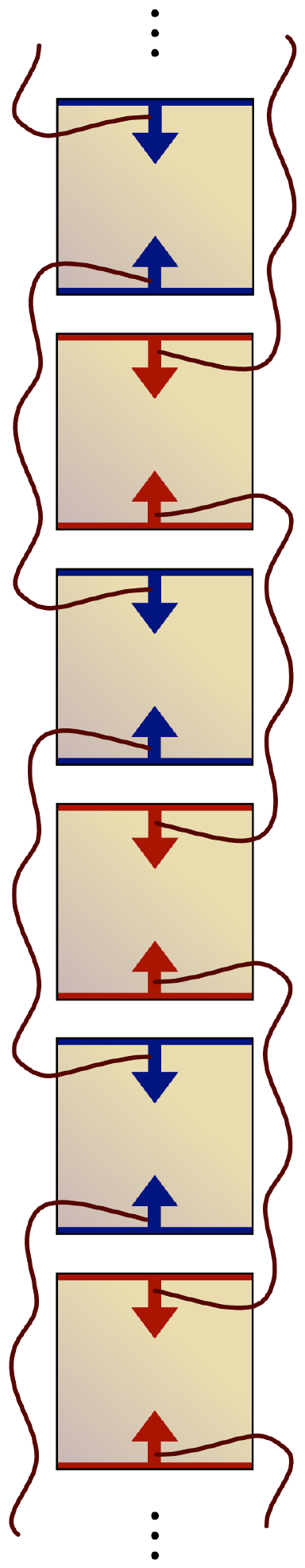, scale=0.35}
\caption{Two independent "lives" lived in parallel by the same particle.}
\end{figure}

\bigskip
\noindent
{\bf Acknowledgements} SP acknowledges support from the European Research Council ERC Advanced Grant NLST, EU grant Q-Essence and the Templeton Foundation.


\begin{thebibliography}{2}
\bibitem{superposition_evolutions} In a completely different context, superpositions of time evolutions were discussed in \cite{time_machine} and experimentally verified in \cite{ernst}.
\end{thebibliography}
\end{document}